\documentstyle[a4, 12pt]{article}
\def\nn{\nonumber}
\def\bea{\begin{eqnarray}}  \def\eea{\end{eqnarray}}
\def\1{{\rm 1\mskip-4.5mu l} }
\newcommand{\beq}{\begin{equation}}
\newcommand{\eequ}{\end{equation}}
\newcommand{\eeq}{\end{equation}}
\newcommand{\noi}{\noindent}

\begin{document}

\begin{center}
\title{\bf Do  $1^- \,c \bar c g$ hybrid mesons exist, do they
 mix  with charmonium ?}  
\author{F. Iddir and S. Safir$^a$ \\[5mm]  
O. P\`ene$^b$}\par
\maketitle
{$^a$ Laboratoire de Physique Th\'eorique \\
Universit\'e d'Oran Es Senia - 
31100   Alg\'erie\footnote{e-mail: iddir@elbahia.cerist.dz}\\
$^b$ Laboratoire de Physique Th\'eorique et Hautes
Energies\footnote{ Laboratoire
associ\'e au
Centre National de la Recherche Scientifique - URA D00063
 }}\\ {Universit\'e
de Paris XI, B\^atiment 211, 91405 Orsay Cedex, France}
\end{center}

\begin{abstract} Hybrid $\,c \bar c g$ of quantum numbers $1^{--}$ are
considered in a quark model with constituant quarks and gluon. The
lowest $J^P=1^-$ states may be built in two ways, $l_g=1$ (gluon
excited) corresponding to an angular momentum between the gluon and the
$c \bar c$ system, while $l_{c\bar c}=1$ (quark excited) corresponds to
an angular momentum between the $c$ and the $\bar c$. The lowest lying
hybrid $J^P=1^-$ state in the flux tube model is similar to the $l_g=1$
in the quark-gluon model. In particular it verifies the selection rule
that it cannot decay into two fundamental mesons.  The $l_{q\bar q}=1$
hybrid may decay into two fundamental mesons, but with decay widths
larger than 1 GeV, which tells that they do not really exist as resonant
states.  Using a chromoharmonic potential, we find no mixing between the
$l_g=1$ and $l_{c\bar c}=1$. More realistic potentials might induce a
strong mixing between them, implying that no hybrid meson exist.  If, on
the contrary, such a strong mixing does not occur, we find, in agreement
with the flux tube model, that only the $l_g=1$ appears as a real
resonant state. In such a case, hybrid mesons may exist as resonances
only if they are decoupled from the ground state channels, which
explains the difficulty to observe them experimentally.
 We
reconsider accordingly the Ono-Close-Page scenario of  mixing between 
charmonium and
charmed-hybrid to explain the anomalies around 4.1 GeV. We  find a very small 
mixing between radially excited charmonium and  hybrid mesons, which forbids 
considering
the $\psi(4.040)$ and $\psi(4.160)$ as combinations of 3S charmonium and 
$l_g=1$ hybrid meson with a large mixing.

\end{abstract}
\begin{flushright} LPT Oran 10/97\\LPTHE Orsay-97-71 \\ hep-ph/9803470
\end{flushright}

\pagestyle{plain}
\section{Introduction}
\hspace*{\parindent}
Beyond the standard meson and baryon states, QCD predicts the existence of 
``exotic'' color singlet states built up of constituant gluons or quarks and 
gluons.
Checking QCD commands to search for exotics.
However exotic hunting has proven along the years to be a very difficult task. 
Some good glueball candidates exist, such as the celebrated $f_0(1500)$ 
\cite{glueball} but it seems very difficult to eliminate all doubts.

This is why the search for exotic non-$q\bar q$ quantum numbers such as
$1^{-+}$ is so important. An undisputable experimental discovery would
mean a real breakthrough.  Such a state would be a very good $q \bar q
g$ candidate. Experimental candidates exist \cite{exotics}, a $1^{-+}$
activity really seems present in $(\eta\pi)_P$ and $(\eta'\pi)_P$
channels \cite{faessler}, but the presence of a real
 resonance is
still questionable.  

 The charmonium mass spectrum presents a noticeable anomaly in the
region of the $\psi(4.040)$ and the $\psi(4.160)$. In short, the
$\psi(4.040)$ is a good candidate for a $3S$ charmonium state, but then
the $\psi(4.160)$ is too close to be a good $4S$ candidate, and its
decay width $\Gamma(\psi(4.160)\to e^+e^-)\simeq \Gamma(\psi(4.040)\to
e^+e^-)$ makes of it a bad $2D$ candidate.  the anomaly has been
tentatively explained by Ono \cite{ono} and more recently by Close and
Page \cite{close} as due to a mixing of the genuine $3S$ charmonium $c
\bar c$ states with a $c \bar c g$ hybrid. We shall refer to this
proposal as the OCP scenario.  According to these authors a $1^{--}$
hidden-charm hybrid should exist in this mass range, and the mixing
angle should be large.

The OCP scenario makes use of a selection rule forbidding the decay of (some) 
hybrids
into  ground state mesons. It results that a $1^{--}$ hidden-charm hybrid of 
mass
$\simeq 4.100$ GeV would be stable if mixing with charmonium did not occur. The
latter selection rule was proven in a quark model with constituant gluon
\cite{tanimoto, nous+ono, iddir, kalashnikova} for the  hybrid with an 
orbitally excited gluon
(sometimes called ``transverse electric''). This selection rule was later 
confirmed
in the flux tube model \cite{isgur, page}. 
The quark model 
also
contains a state in which the gluon is in an S-wave but the $c$ and $\bar c$ 
quarks
are in a relative P-wave. The latter quark-excited hybrids do not present the
above-mentioned selection rule. 

Indeed the experimental search for hybrids needs some theoretical
understanding of the expected signatures of such a state. Unhappily, no
QCD-based method can be operative in such a complex phenomenon. Lattice
QCD may say a word about hybrid masses \cite{lattice}, but certainly not
about decay. To our knowledge, besides some old work using QCD surm
rules \cite{QCDSR}, the flux tube model and the quark model with
constituant gluon are the two main tools used to study hybrid mass
spectrum and decay.

The relations between these two models deserves some theoretical
study. We will try to initiate this.  Since we need to have some idea
about the dynamics we will solve an oversimplified chromo-harmonic
Hamiltonian \cite{gavela}.  This will also provide us with the wave
functions of the hybrids and of the final mesons.

We will then compute the decay widths of both type of hybrids into
ground state mesons according to the quark-gluon constituant model
\cite{tanimoto, nous+ono, iddir, kalashnikova}. The gluon-excited one
does not decay as recalled above. The width of the quark-excited one
will turn out to be exceedingly large. This excludes the latter for the
OCP scenario. The consequences will be discussed.

 The OCP scenario could then survive with the gluon-excited
 hybrid {\it if the transition matrix elements between $3S$ $c \bar c$
and the hybrid allow for a large mixing angle}.  Unhappily the mixing
will turn out much too small.
 
 In section \ref{model}, we present and solve the model: spectrum and 
 decay widths. 
In section \ref{decay} we compute the $1^{--} c\bar c g$ hybrid decay into
ground state mesons and derive the selection rule for the gluon-excited one.
In section \ref{OCP} we compute the mixing of the  $1^{--} c\bar c g$
hybrids with the $3S c \bar c$ charmonium state.
We then conclude.

\section{The Quark Model with constituant gluon} \label{model}
\hspace*{\parindent} We shall now describe the Hamiltonian, the wave
functions and the decay of hybrid mesons. The hybrid meson theoretical
study is still in its enfancy, although it started some time ago,
presumably because no such meson has yet been unambiguously observed. It
is therefore premature to recourse to sophisticated tools which anyhow
would demand a number of arbitrary inputs. We need some robust method
able to answer to a series of questions: spectrum, decay, mixing, etc.
We choose the non-relativistic constituant quark model, well known to be
semi-quantitqtively successful in describing the hadron spectrum, and
will compare its prediction to the more extensively used flux tube
model. To make our life even simpler we will use a chromo-harmonic
potential very easy to diagonalize.
 We hope to catch by this method the gross features of hybrid physics.
 This extreme simplification may nevertheless introduce some artifact as we
 shall discuss later on.

\subsection{A simple chromo-harmonic Hamiltonian  and the wave
functions of the hybrid and the final mesons} 
\hspace*{\parindent}

We start from the Hamiltonian:

\beq
H = {p^2_q \over 2m_q} + {p_{\bar{q}}^2 \over 2m_{\bar{q}}} + {p_g^2 \over 2
m_g} + V\left (r_q, r_{\bar{q}}, r_g \right )
\label{1e} 
\eeq

\noi where~: 
\bea
V \left ( r_q, r_{\bar{q}}, r_g \right ) &=& b_0 \sum_{a}
\Big [ T_q^a \ T^a_{\bar{q}} (r_q - r_{\bar{q}})^2 + T_g^a \ T^a_{\bar{q}} 
(r_g - r_{\bar{q}})^2 \nn \\
&&+ T_g^a \ T_q^a(r_g - r_q)^2 \Big ] \quad .
\label{2e}
\eea

\noi The potential (\ref{2e}) is spin independent
(nonrelativistic
quark model), it is acting on color and configuration space. It represents a
matrix in color space.

We assume in our model an harmonic interaction term between the constituants 
which
is proportional to the color matrix $(T_j^a)$, this choice, inspired from
gluon exchange in perturbative QCD, expresses the idea that the confining
interaction between two constituant must be proportionial to their color charge
\cite{gavela}.
 
Note that~:

\beq
\label{3e}
\sum_a T_i^a \ T_j^a = {1 \over 2} \left [ \left ( T_i^a + T_j^a \right )^2 - 
\left
( T_i^a \right )^2 - \left ( T_j^a \right ) \right ]^2
 \eeq

\noi where $T^a$ are the generators fo the SU(3) group in the different
representations of the constituants, example~:

\beq \label{4e} \left ( T_q^a \right )_{ij} = {\lambda_{ij}^a \over 2}
\quad , \quad \left ( T_{\bar{q}}^a \right )_{ij} = {\lambda^{a*}_{ij}
\over 2} \quad \hbox{and} \quad \left ( T_g^a \right )_{bc} = - i
f_{bc}^a = F_{bc}^a \ .  \eeq

\noi $i + j$ is necessarily in a well defined color representation, 
such
that combined with the third constituant it gives an overall color singlet. 
Hence
not only $T_i^{a^2}$ and $T_j^{a^2}$ are the identity times
 eigenvalues of the Casimir operator but 
also $(T_i^a +
T_j^a)^2$ is the Casimir eigenvalue for the color representation of the third 
constituant.

Finally~:

\beq
\sum_a T_i^a \ T_j^a = \alpha_{ij} \ \1
\label{5e}
\eeq

\noi where $\1$ is identity matrix in color space and 

\beq
\label{6e}
\alpha_{ij} \equiv {1 \over 2} \left [ C(k) - C(i) - C(j) \right ]
\eeq

\noi where $k$ is the third constituant and $C(i)$ is the Casimir of $i$.

~From $C(3)=4/3$ and $C(8)=3$ we find 

\bea
\label{7e}
&&\alpha_{q \bar{q}} = 1/6 \nn \\
&&\alpha_{qg} = \alpha_{\bar{q}g} = - 3/2 \quad .
\eea

\noi Finally the potential $V$ becomes independent of colors indices and all
information about color  representation are encoded in the number 
$\alpha_{ij}$~:

\beq
\label{8e}
V \left ( r_q , r_{\bar{q}}, r_g \right ) = b_0 \sum_{{i,j=q,\bar{q}, g
\atop i < j}} \alpha_{ij} (r_i - r_j )^2
 \eeq

~From (\ref{1e}), (\ref{2e}) and (\ref{3e}) we obtain the Hamiltonian for
color singlet $q\bar{q}g$ system~:

\beq
H_{hyb} = {P^2 \over 2M} + {p_{q\bar{q}}^2 \over 2 \mu_{q\bar{q}}} + {k^2 
\over 2
\mu_{g}} - {7b_0 \over 12} r_{q\bar{q}}^2 - 3 b_0 r^2
 \label{9e}
\eeq

\noi where $P = p_q + p_{\bar{q}} + p_g$ is the center of mass momentum of
$q\bar{q}g$ system, $p_{q\bar{q}} = \displaystyle{{m_{\bar{q}} p_q - m_q 
p_{\bar{q}}
\over m_q + m_{\bar{q}}}}$ is the  relative momentum between $q$ 
and
$\bar{q}$, $k = \displaystyle{{p_g (m_q + m_{\bar{q}}) - (p_q +
p_{\bar{q}}) m_g \over m_g (m_q + m_{\bar{q}})}}$ is the 
relative momentum between $g$ and $q\bar{q}$. Their conjugate variables are 
$r_{q\bar{q}} = r_q - r_{\bar{q}}$ and
$ r = r_g - \displaystyle{{r_q + r_{\bar{q}} \over 2}}$.
\beq
M = m_q + m_{\bar{q}} + m_g \qquad ; \quad \mu_{q\bar{q}} = {m_q m_{\bar{q}} 
\over
m_q + m_{\bar{q}}} \quad \hbox{and} \quad \mu_g = {(m_q + m_{\bar{q}})
m_g \over m_q + m_{\bar{q}} + m_g} \quad . 
\label{10e}
\eeq 

\noi The eigenstates of the Schr\"odinger equation (\ref{9e}) are~: 

\beq \psi_{\ell_{q\bar{q}}}^{m_{q\bar{q}}}(p_{q\bar{q}}) = \left \{ {16
\pi^3 R_{q\bar{q}}^{2\ell_{q\bar{q}} + 3} \over \Gamma \left ( {3 \over
2} + \ell_{q\bar{q}} \right )} \right \}^{1/2} \
y_{\ell_{q\bar{q}}}^{m_{q\bar{q}}}(\vec{p_{q\bar{q}}}) \ e^{- {1 \over
2} \left ( R_{q\bar{q}}^2 p_{q\bar{q}}^2 \right )} \label{11e} \eeq

\noi and

\beq
\psi_{\ell_g}^{m_g}(k) = \left \{ {16 \pi^3 R_g^{2\ell_g + 3} \over \Gamma
\left ( {3 \over 2} + \ell_g \right )} \right \}^{1/2} \ 
y_{\ell_g}^{m_g}(\vec{k}) \
e^{- {1 \over 2} R_g^2 k^2} \label{12e} \eeq

\noi where

\beq
y_{\ell}^m (\vec{p}) \equiv p^{\ell} Y_{\ell}^m (\theta, \Omega ) \quad .
\label{13e}
\eeq

\noi and where :
\begin{eqnarray}
\label{14b}
R_{q\bar{q}}^2 & = & (2 \mu_{q\bar{q}}(-7b_0/12))^{-1/2} \nn \\
R_{g}^2 & =& (2 \mu_{g}(-3 b_0))^{-1/2}
\end{eqnarray}

It is noticeable that in the case of the chromoharmonic Hamiltonian 
(\ref{1e})-(\ref{2e}) the quantum numbers $l_{q\bar q}$ and $l_g$ are 
diagonalized. This feature would not remain valid for a more general potential.
This will be discussed later. 

\noi Analogously the Hamiltonian (\ref{1e})-(\ref{2e}) restricted to a meson 
state writes~:

\beq
H_{meson} = {P^2 \over 2M} + {p^2 \over 2 \mu}  - {4b_0 \over 3} r^2 
\label{9b}
\eeq

\noi where $P = p_q + p_{\bar{q}}$,  $p = \displaystyle{{m_{\bar{q}} p_q - m_q 
p_{\bar{q}}
\over m_q + m_{\bar{q}}}}$, conjugate to  $r = r_q - r_{\bar{q}}$ and where~:

\beq
M = m_q + m_{\bar{q}} \qquad ; \quad \mu = {m_q m_{\bar{q}} \over
m_q + m_{\bar{q}}} \quad . 
\label{10b}
\eeq 

\noi The final states mesons, which will be labelled $B$ and $C$, will be
described by teh eigenstates of eq. (\ref{9b}):

\beq
\psi_{\ell_B}^{m_B}(p) = \left \{ {16 \pi^3  R_B ^{2\ell_B + 3} \over \Gamma 
\left ({3 \over 2}+ \ell_B \right )}
\right \}^{1/2} \ y_{\ell_B}^{m_B}(\vec{p})  \ e^{- {1 \over 2} R_B^2 p^2} 
\label{33e} \eeq

\noi with :
\beq
R_{B}^2 =  (2 \mu_{B} (-4b_0/3))^{-1/2}
\label{11b}
\eeq

\noi  and the same for $\psi_C$.  

 \subsection{General
formulae for hybrid decay into two mesons} \hspace*{\parindent} The
Classification and the decay of hybrid mesons in a constituant model
have been studied in \cite{tanimoto, nous+ono, iddir, kalashnikova}~; we
will use here the notations of \cite{nous+ono}~: \\

$\ell_g$ : is the relative orbital momentum between the gluon and the
$q\bar{q}$ center of mass \par
$\ell_{q\bar{q}}$ : is the relative orbital momentum between $q$ and
$\bar{q}$ \par
$S_{q\bar{q}}$ : is the total quark spin \par
$j_g$ : is the total gluon angular momentum \par
$L$~: $\ell_{q\bar{q}} + J_g$. \\

\noi The parity and charge conjugation of the hybrid are given by~:

\begin{eqnarray}
\label{14e}
&&P = (-)^{\ell_{q\bar{q}} + \ell_g} \nn \\
&&C = (-)^{\ell_{q\bar{q}} + S_{q\bar{q}} + 1} \ \ \ .
\end{eqnarray}

\noi To lowest order the decay is described by the matrix element of the QCD:

\beq
H = g \int dx \ \Psi (x) \ \gamma_{\mu} {\lambda^a \over 2} \psi (x) \
A_a^{\mu}(x)  \label{15e}
\eeq

\noi we expand at $t = 0$

\beq
\psi (x) = \sum_{s=1}^2 \int {d^3p \over (2 \pi )^3} e^{i \vec p \vec
 x} \left ( u_{ps} \
b_{ps} + v_{ps} \ d^+_{ps} \right )  \label{16e}
\eeq

\beq
\label{17e}
A_a^{\mu} (x) = \sum_{\lambda = 1}^2 \int {d^3k \over \sqrt{2w} (2 \pi)^3}
 \ \varepsilon_{\kappa \lambda a}^{\mu} \left ( a_{\kappa \lambda a} \
e^{i\vec k\vec x} + a_{\kappa \lambda a}^+ \ e^{-i\vec k\vec x} \right )  \eeq

\noindent with $\varphi_a$ being color representation and~:

\begin{eqnarray}
\label{18e}
&&\left \{ b_{ps\rho} , b^+_{p's'\rho'} \right \} = (2 \pi)^3 \ \delta^3 (p - 
p') \
\delta_{s,s'}\ \delta_{\rho ,\rho'} \nn \\
&&\left [ a_{\kappa \lambda a} , a^+_{\kappa ' \lambda ' a'} \right ] = (2 \pi 
)^3 \
\delta^3 (k - k') \ \delta_{\lambda ,\lambda '}\ \delta_{a ,a '} \quad . 
\end{eqnarray}

\noi The hamiltonian annihilating a gluon and creating a quark pair is~:

\beq \label{19e} H = g \sum_{ss'\lambda\zeta\zeta' } \int {d^3p \ d^3k \
d^3p' \over \sqrt{2\omega} (2 \pi )^9} (2 \pi )^3 \ \delta_3 (p - p' -
k) \left(\bar{u}_{ps} \ \ \gamma_{\mu} {\lambda^a_{\zeta\zeta' }\over 2}
v_{p's'}\right)\ b^+_{ps\zeta} \ d^+_{p's'\zeta'} \ \
\varepsilon_{\kappa \lambda }^{\mu} \ a_{\kappa \lambda a} \eeq

\noi in the nonrelativistic limit~:

\beq
\bar{u}_{ps} \ \gamma_{\mu} \ v_{p's'} \ \varepsilon_{\kappa \lambda}^{\mu} =
\chi_s^+ \vec \sigma \ \chi_{s'} \vec \varepsilon_{\kappa\lambda} \label{20e}
\eeq

\noindent where $\chi_{s'}$ is the antiquark spinor in the complex conjugate
representation.

The meson state and the hybrid state are given respectively in the
nonrelativistic approximation by~:

\begin{eqnarray} 
|B,M_B> &=& \sum_{ss'\atop am \mu} \int {d^3p_q \
d^3p_{\bar{q}} \over (2 \pi)^6 \sqrt{3}} \ \chi_{ss'}^{\mu_B} (2 \pi)^3
\delta_3 \left ( p_B - p_q - p_{\bar{q}} \right ) \\ &&
\psi_{\ell_B}^{m_B}\left ( {m_{\bar{q}} p_q - m_q p_{\bar{q}} \over m_q
+ m_{\bar{q}}} \right ) <\ell_B \ m_B \ S_B \ \mu_B | J_B \ M_B>
b^+_{p_q sa} \ d^+_{p_{\bar{q}}s'a}|O> \quad . \nn \label{21e}
\end{eqnarray}

\noi with $\int d^3p/(2 \pi)^3 |\psi_B (p)|^2 = 1$ and analogously for the $C$.

Note here that $q$ and $\bar{q}$ can be different in flavour. The hybrid meson
state in its rest frame is given by

\begin{eqnarray}
|A,L,M> &=& \sum \int {d^3p_q \ d^3p_{\bar{q}} d^3k \over (2 \pi)^9} \
\chi_{s_qs_{\bar{q}}}^{\mu_{q\bar q}} (2 \pi )^3 \ \delta_3 (k + p_q +
p_{\bar{q}}) {\lambda^{c_g}_{c_q \ c_{\bar{q}}} \over 4}\\
&&\psi_{\ell_{q\bar{q}}}^{m_{q\bar{q}}} \left ( {m_{\bar{q}} p_q - m_q 
p_{\bar{q}} \over m_q +
m_{\bar{q}}} \right ) \ \psi_{\ell_g}^{m_g}\left ( {(m_q + m_{\bar{q}})k - m_g 
(p_q +
p_{\bar{q}}) \over m_q + m_{\bar{q}} + m_g} \right ) \nn \\  
&&<\ell_g \ m_g \ 1 \ \mu_g |J_g M_g> <\ell_{q\bar{q}} \ m_{q \bar{q}} \ J_g \ 
M_g
|L m'> <\ L \ m' \ S_{q \bar{q}} \ \mu_{q \bar{q}}|JM > \nn \\ 
&&b^+_{p_qs_qc_q} \ d^+_{p_{\bar{q}}s_{\bar{q}}c_{\bar{q}}} \
 a^+_{\kappa \mu_g c_g}
|0>  \nn
\label{22e}
\end{eqnarray}

\noi where $c_q$, $c_{\bar{q}}$ and $c_g$ are the color charge of the quark,
antiquark and gluon with $c_q = 1, \cdots 3$~; $c_{\bar{q}} = 1, \cdots 3$~; 
$c_g
= 1, \cdots 8$ and a sum over repeated indices is understood. Using (\ref{19e}) 
one gets in a straightforward manner the matrix
element $<B,M_B \ C,M_C|H|A,L,M>$ between an hybrid state $A$ and two mesons $B$ 
and $C$~:

\beq
<B,M_B \ C,M_C|H|A,L,M> = gf (A, B, C) (2 \pi)^3 \ \delta_3(p_A - p_B - p_C)
\label{23e}
\eeq

\noi where $f(A,B,C)$ representing the decay amplitude by~:

\begin{eqnarray} f(A, B, C) &=& \sum_{m_{q\bar{q}}, m_g, m_B, m_C \atop
\mu_{q\bar{q}}, \mu_g, \mu_B, \mu_C} \ \Omega \phi \ X(\mu_{q \bar{q}},
\mu_g ; \mu_B , \mu_C ) \\ &&I \left ( m_{q\bar{q}}, m_g ; m_B, m_C, m
\right ) <\ell_g \ m_g \ 1 \ \mu_g |J_g M_g> \nn \\ &&<\ell_{q\bar{q}} \
m_{q\bar{q}} \ J_g \ M_g |L m'> <L \ m' \ S_{q\bar{q}} \ \mu_{q\bar{q}}
|J M> \nn \\ &&< \ell_B \ m_B \ S_B \ \mu_B |J_B M_B> < \ell_C \ m_C \
S_C \ \mu_C |J_C M_C> \nn \label{24e} \end{eqnarray}

\noi where $\Omega$, $X$, $I$ and $\phi$ are the color, spin, spatial and
flavour overlaps. $\Omega$ is given by~:

\beq
\Omega = {1 \over 24} \sum_a {\rm Tr} \ (\lambda^a)^2 = {2 \over 3}
\label{25e}
\eeq

\noi from :
\beq
\chi_{\mu_1}^+ \ \sigma^{\lambda} \ \ \chi_{\mu_2} 
= \sqrt{3} < {1 \over 2} \ \mu_2 \ 1 \ \lambda |{1 \over 2} \ \mu_1 >  
\label{26e}
\eeq

\noi where $\sigma^{\pm 1}\equiv \mp{1 \over \sqrt{2}}(\sigma_x\pm i \sigma_y)$ 
and 
$\sigma^0=\sigma_z$,
 we obtain the spin overlap~: 

\begin{eqnarray}
X(\mu_{q\bar{q}}, \mu_g; \mu_B, \mu_C) &=& \sum_s \sqrt{2} \left [
\begin{array}{lll} 1/2 &1/2 &S_B \cr 1/2 &1/2 &S_C \cr S_{q \bar{q}} &1 &S
\cr \end{array} \right ] <S_{q\bar{q}} \ \mu_{q\bar{q}} \ 1 \ \mu_g |S \ \mu_{q
\bar{q}} + \mu_g> \nn \\
&&<S_B \ \mu_B \ S_C \ \mu_c |S \ \mu_B + \mu_C > 
\label{27e}
\end{eqnarray}

\noi where :

\beq
\left [ \begin{array}{lll} 1/2 &1/2 &S_B \cr 1/2 &1/2 &S_C \\ S_{q
\bar{q}} &1 &S \end{array}\right ] = \sqrt{(2 S_B + 1) (2S_C + 1) 3(2S_{q
\bar{q}} + 1)} \left \{ \begin{array}{lll} 1/2 &1/2 &S_B \cr 1/2 &1/2 &S_C
\cr S_{q\bar{q}} &1 &S \end{array} \right \}
\label{28e}
\eeq

\noi the spatial overlap is given by:

\begin{eqnarray}
\label{29e}
&&I \left ( m_{q \bar{q}} , m_g; m_B, m_C, m \right ) = \int \int {dp \ dk \over
\sqrt{2\omega} (2 \pi)^6} \psi_{\ell_{q\bar{q}}}^{m_{q\bar{q}}} (p_B - p) \
\psi_{\ell_g}^{m_g} (k) \\
&&\psi_{\ell_B}^{m_B^*} \left ( {p_B m_{\bar{q}_i} \over m_q + m_{\bar{q}_i}} -
p - {k \over 2} \right ) \psi_{\ell_C}^{m_C^*} \left ( -{m_{q_i} p_B \over
m_{q_i} + m_{\bar{q}}} + p - {k \over 2} \right ) d\Omega_B \nn
Y_{\ell}^{m^*}(\Omega_B) \end{eqnarray}

\noi The $Y_{\ell}^{m^*}(\Omega_B)$
spherical harmonic, integrated over the $4\pi$ solid angle for  $\Omega_B$
projects out the decay amplitude on the orbital momentum $\ell, m$ between 
the two final mesons. The 
and ${q_i
\bar{q}_i}$ form  the created  quark pair. 

Finally:

\beq
\phi = \left [ \begin{array}{lll} i_1 &i_3 &I_B \cr i_2 &i_4
&I_C \cr I_A &0 &I_A \cr \end{array}\right ] \eta \varepsilon  \label{30e}
\eeq

\noi where the $I$'s ($i$'s) label the hadron (quark) isospins, $\eta = 1$ if
the gluon goes into strange quarks and $\eta = \sqrt{2}$ if it goes into
nonstrange ones. $\varepsilon$ is the number of diagrams con\-tri\-bu\-ting to 
the
decay. Indeed one can check that since $P$ and $C$ are conserved, two diagrams
contribute with the same sign and magnitude for allowed decays while they
cancel for forbidden ones. In the case of two identical final particles,
$\varepsilon = \sqrt{2}$. The partial width is then given by~:

\beq
\Gamma (A \to BC ) = 4 \ \alpha_s \ |f(A, B, C)|^2 {P_B \ E_B \ E_C \over M_A}
\label{31e}
\eeq

\noi with~:

\begin{eqnarray}
&&P_B^2 = { \left [ M_A^2 - (m_B + m_C)^2 \right ] \left [ M_A^2 - (m_B - m_C)^2
\right ] \over 4M_A^2} \qquad ;\nn \\
&&E_B = \sqrt{P_B^2 + m_B^2} \qquad , \quad E_C = \sqrt{P_B^2 + m^2_C} \quad .
\label{32e}
\end{eqnarray}

\section{The $1^{--}$ hybrids decay and the selection rule}
\label{decay}
\hspace*{\parindent}

\subsection{Classification of the $1^{--}$ hybrid states}

Let us now consider the lightest $J^{PC}=1^{--}$ hybrid mesons, using
the quark model with a constituant gluon \cite{tanimoto, nous+ono,iddir,
kalashnikova} as developed in section 2.
 Eq. (\ref{14e}) implies $\ell_{q\bar{q}} = S_{q\bar{q}}$ and 
$\ell_{q\bar{q}}+l_g 
$ odd. The lightest such states are $\ell_{q\bar{q}} =S_{q\bar{q}} =0, 
\ell_g =1$, which we shall refer to as the gluon-excited hybrid, and 
$\ell_{q\bar{q}} =S_{q\bar{q}} =1, 
\ell_g =0$, which we shall refer to as the quark-excited hybrid.
In the case of the gluon-excited one (respectively the quark excited one)
 we obtain from 
the Clebsch Gordan of the eq. (\ref{21e}) that $L =J =J_g =1$ 
(respectively $L=0,1,2; J_g =1$). 

\begin{table}[htb]
\begin{center}
\begin{tabular}{|c|c|c|c|c|c|c|c|}
\hline
& & & & & & &\\
$P$ & $C$ & $\ell_g$ & $\ell_{q\bar{q}}$ & $J_g$ &$S_{q\bar{q}}$ &$L$ & $J$\\ 
& & & & & & &\\
\hline
$-$ & $-$ & 0 & 1 & 1 &1 &0 &1 \\ 
$-$ & $-$ & 0 &1 &1 &1 &1 &1 \\ 
$-$ & $-$ & 0 &1 &1 &1 &2 &1\\ 
$-$ & $-$ & 1 &0 &1 &0 &1 &1\\ 
\hline
\end{tabular}
\label{tab1}\vskip 5mm
\caption{\leftskip 1pc \rightskip 1pc
\baselineskip=10pt plus 2pt minus 0pt Lowest $J^{PC}=1^{--}$ 
 $q\bar{q}g$ hybrid mesons and their quantum numbers}
\end{center}
\end{table}

\noi Notice that the gluon-excited state, verifying $\ell_g =J_g =1$, is
one member of what has been sometimes called a ``tranverse-electric''
hybrid \cite{nous+ono}, while the quark-excited states with $\ell_g\ne
J_g$ are ``transverse-magnetic''. We will not use this nomenclature
which does not seem too convenient.

\subsection{Decay into ground state mesons and the selection rule}

Let us compute the integral  (\ref{29e}) for ground $l_B=l_C=0$ assuming
$R_B=R_C$:

\begin{eqnarray} &&I(m_{q \bar{q}}, 0; 0, 0, m) = 2^4 \sqrt{{\pi \over 3
\omega}} {R_{q \bar{q}}^{3/2+\ell_{q\bar{q}}} \ R_g^{3/2+\ell_g} \ R_B^5
\over \left ( R_g^2 + R_B^2/2 \right )^{3/2} \left ( R_{q \bar{q}}^2 +
2R_B^2 \right )^{5/2}} {2m_q \over m_q + m_{\bar q_i}} \ P_B \ \\ &&
\exp \left\{- {P_B^2 \over 2} \left[ R_{q \bar{q}}^2 + {2
m^2_{\bar{q}_i}R_B^2 \over (m_q + m_{\bar{q}_i})^2}
 - { \left [ 2m_{q_i} \ R_B^2 + \left ( m_q + m_{\bar{q}_i} \right )
R_{q\bar{q}}^2 \right ]^2 \over \left ( R^2_{q \bar{q}} + 2R_B^2 \right ) \left
( m_q + m_{\bar{q}_i} \right )^2} \right]\right\} \
\delta_{\ell_{g,0}}\delta_{\ell_{q \bar{q}}, \ell} \
\delta_{m_{q \bar{q}}, m}\nn
\label{36e}
 \end{eqnarray}

\noi the $\delta_{\ell_{g,0}}$ term in the last formula tells us that a 
gluon-excited hybrid cannot decay
into two ground state mesons. This is the well known 
{\it  selection rule for the gluon-excited hybrids decay} \cite{tanimoto,
 nous+ono,iddir, isgur, page}. This selection rule is a basic ingredient 
 of the OCP scenario as will be discussed below.
 The $\delta_{\ell_{q \bar{q}}, \ell}$ factor shows us 
that the intermeson orbital momentum is a direct
measure of the interquark orbital momentum in the hybrid when $R_B=R_C$.  

\subsection{Decay widths of quark-excited hybrids}

To fix the $b_0$ parameter of the harmonic oscillator potential (\ref{9e})
we will use the 
  level spacing of the charmonium, charmed and the light
mesons with Hamiltonian (\ref{9b}).
We present in this table the different values of $b_0$~:
\vskip 5mm
\begin{table}[htb]
\begin{center}
\begin{tabular}{|l|l|l|l|l|}
\hline
& & & & \\
meson sector &$\omega_{exp}$(GeV) &$\mu$(GeV) &$R^2_{exp}$(GeV$^{-2}$) 
&$b_0$(GeV$^3$)
\\ & & & & \\
\hline 
 light meson &\quad 0.465 &$\ $ 0.175 &\qquad 12.28 &$-$ 0.015 \\
\hline
 charmonium meson &\quad 0.295 &$\ $ 0.85 &\qquad $\ $ 3.98 &$-$ 0.03 \\
\hline
charm meson &\quad 0.453 &$\ $ 0.3 &\qquad $\ $ 7.35 &$-$ 0.025 \\
\hline
\end{tabular}
\end{center}
\label{tab2}\vskip 5mm
\caption{\leftskip 1pc \rightskip 1pc
\baselineskip=10pt plus 2pt minus 0pt $b_0$ fitted from meson level spacing.
$\omega_{exp}$ is the roughly averaged experimental level spacing
 between the ground state and 
the first excited state and $\mu$ is the reduced mass in eq. (\ref{10b}). }
\end{table}
\vskip 5mm

The fact that in table 2 the value of $b_0$ varies from line to line
reflects the well known fact that the harmonic oscillator potential is
not a good potential for mesons. However, we have checked that none of
our results in this paper will depend drastically on the variation of
$b_0$ within the range $-0.015$ to $-0.03$ GeV $^3$. It is why we will
only present the results for one central value $b_0 = -0.02$ GeV$^3$ for
decays widths and the mixing.

In Eq. (\ref{14b}) we use the following masses~: 

\[ m_c = m_{\bar{c}} = 1.7 \ \hbox{GeV} ,\quad m_u = m_d = 0.35 \
\hbox{GeV} \quad \hbox{and} \quad m_g = 0.8 \ \hbox{GeV} \quad . \]

\noindent From (\ref{14b}-\ref{11b}) we obtain the corresponding radii with  
$b_0 = - 0.02$ GeV$^3$:
 \begin{eqnarray}
&&R_{c\bar{c}}^2 = 7.1 \ {\rm GeV}^{-2} \\
&&R_B^2 = 7.9 \ {\rm GeV}^{-2} \nn \\
&&R_g^2 = 3.58 \ {\rm GeV}^{-2} \nn  \quad .
\label{39e} 
\end{eqnarray}  
\noindent Remeber that the level spacing is given by $\omega = 1/(\mu R^2)$.

\noindent The decays width of the hybrid Charmonium $1^{--}$~:  

\vskip 5mm
\begin{table}[htb]
\begin{center}
\hspace*{-1.cm}
\begin{tabular}{|c|c|c|c|c|c|c|c|c|c|c|}
\cline{7-10} 
\multicolumn{6}{c}{} &\multicolumn{4}{|c|}{$\Gamma_{D^{*0}\bar{D}^{*0}} =
\Gamma_{D^{*-}{D}^{*+}}$} &\multicolumn{1}{c}{}  \\ 
\hline
& & & & & & & & & &  \\ 
$L$ &$\Gamma_{D^0\bar{D}^0}$ &$\Gamma_{D^+D^-}$ &$\Gamma_{D^+_s D_s^-}$
&$\Gamma_{D^{*0}\bar{D}^0}$ &$\Gamma_{\bar{D}^{*0}D^0}$   
 &$S=0$ &$S=1$ &$S=2$ &$\Gamma_{tot\ D^{*0}\bar{D}^{*0}}$
&$\Gamma_{tot}(4.04)$ \\ 
\hline 
0 &176 &179 &135 &357 &357 &6,7 &0 &135 &141,7 &1487,1 \\
\hline 
1 &528 &536 &405 &268 &268 &20 &0 &101,5 &121,5  &2248 \\
\hline
2 &880 &894 &674 &446 &446 &34 &0 &6,7 &40,7  &3421,1 \\
\hline
\end{tabular}
\end{center}
\caption{\leftskip 1pc \rightskip 1pc
\baselineskip=10pt plus 2pt minus 0pt {\sl Predicted widths in
$\alpha_s$ Mev for the decay of a quark-excited ($l_{c\bar c}=1$) 
$c\bar c g\, 1^{--}$ hybrid meson of mass $(4.04)$ GeV} }
\label{tab3}\vskip 5mm
\end{table}
\vskip 5 truemm

\begin{table}[htb]
\hspace*{-1.cm}
\begin{tabular}{|c|c|c|c|c|c|c|c|c|c|c|}
\cline{7-10} 
\multicolumn{6}{c}{} 
&\multicolumn{4}{|c|}{$\Gamma_{D^{*0}\bar{D}^{*0}} = \Gamma_{D^{*-}{D}^{*+}}$} 
&\multicolumn{1}{c}{}  \\ 
\hline
& & & & & & & & & & \\ 
$L$ &$\Gamma_{D^0\bar{D}^0}$ &$\Gamma_{D^+D^-}$ &$\Gamma_{D^+_s D_s^-}$
&$\Gamma_{D^{*0}\bar{D}^0}$ &$\Gamma_{\bar{D}^{*0}D^0}$ 
&$S=0$ &$S=1$ &$S=2$ &$\Gamma_ {tot\ D^{*0}\bar{D}^{*0}}$
&$\Gamma_{tot}(4.16)$ \\ 
\hline 
0 &132 &136 &196 &372 &372  &57,5 &0 &1151 &1208,5 &3625 
\\
\hline 
1 &396 &407,5 &588 &279 &279  &173 &0 &863,5 &1036,5 
&4022,5 \\
\hline
2 &659 &679 &980 &465 &465  &288 &0 &57,5 &345,5 &3939 \\
\hline
\end{tabular}
\vskip 1.5 cm
\caption{\leftskip 1pc \rightskip 1pc
\baselineskip=10pt plus 2pt minus 0pt {\sl Predicted widths in
$\alpha_s$ Mev quark-excited ($l_{c\bar c}=1$) 
$c\bar c g 1^{--}$ hybrid meson of mass {$(4.16)$ GeV}}}
\label{tab4}\vskip 5mm
\end{table}

\noi The decay widths of the quark-excited hybrid charmonium are 
very large, much larger
than the level spacing, meaning that such states do not really emerge from the 
continuum of the two
meson spectrum. In other words, the quark-excited hybrids do decay before 
they have had time to 
really exist as a
hadron. {\it The quark-excited hybrid do not exist as resonances}.

At this stage, in the chromo-harmonic model we have a very strongly cut
distinction between the gluon excited ($\ell_g = 1$) hybrids which are
not allowed to decay, and the quark excited ones $(\ell_{q\bar{q}} =
1$), too broad to be considered as resonances. {\it We may fear} that
using a more realistic potential, the mixing between these two types of
hybrids will produce only broad eigenstates, i.e. lead us to conclude
{\it that maybe $1^{--}$ hybrids do not really exist}.

\subsection{A comparative look at the flux tube model}
\hspace*{\parindent}

 The flux tube model
\cite{paton} assumes that hybrids are predominantly quark-antiquark
states moving on an adiabatic surface generated by an excited color flux tube. 
There is no constituant gluon as in the quark-gluon constituant model, the gluon
degrees of freedom are treated as collective excitations of the color flux. In 
practice the color is treated as a string, and the excited states are  
represented by the excited modes of the string. 

This picture seems to be close to the picture of a gluon orbitally
excited around the center of mass. Indeed, looking at eq. (4) in
\cite{isgur}, the wave function of the excited hybrid in the flux tube
model is very similar to the one in eq. (\ref{12e}) in this
paper\footnote{But it is totally different from the quark-excited state
eq. (\ref{11e}).}. The variable $y$ in \cite{isgur} happens to be the
variable $k$ in this paper.  It's why the flux tube model predicts the
same selection rule as the Quark model with constituant gluon in the
gluon-excited mode.

The quark excited mode looks similar to the next to lowest lying hybrid
in the flux tube model\footnote{We are very much indebted to Philip Page
for having drawn our attention on this similarity.} formed from a
vanishing total orbital momentum in the hybrid and $S_{c\bar c} =
1$. Indeed, the first line in table 1 has exactly the latter quantum
numbers. In the flux tube model, the state with $S_{c \bar c} =1$ does
not follow the selection rule for the decay into ground state mesons
\cite{page}. One difference between the two models is that in the
constituant model the gluon excited mode is lighter than the quark
excited, as can be seen from eq. (\ref{9e}): the harmonic coefficient of
the gluon mode is $-3 b_0$, much larger than that of the quark mode $-7
b_0/12$, while the reduced masses are of the same order. Using the
parameters of the preceding section, we find an excitation energy of
$\sim$ 200 MeV for the quark excited mode, versus $\sim$ 600 MeV for the
gluon excited one.

\section{Hybrid-charmonium mixing} \label{OCP} \hspace*{\parindent} In
the charmonium spectroscopy, the $\psi (4040)$ and $\psi(4160)$ ($J^{PC}
= 1^{--}$) were traditionally believed and regarded as $3S$ and $4S$
$c\bar{c}$ respectively, but the fact that these two states are only
split by about $\sim$~100~MeV while the normal splitting in the
conventional charmonium picture is about $\sim$~400~MeV has lead to ask
some questions about the real composition of these objects.

A suggestion to explain the anomalous narrow $\psi(4040)-\psi(4160)$ splitting 
is to assume the
existence of an additional state: an hybrid meson with a mass around 4.1 GeV
mixed with conventional $\psi (3S)$ charmonium around 4.1 GeV. 
The Ono-Close-Page (OCP) scenario \cite{ono}, \cite{close} 
 propose the physical eigenstates to be~:

\beq
\psi_{\pm} \equiv {1 \over \sqrt{2}} \left ( \psi(3S) \pm H_C \right )
\label{37e}
\eeq

\noi which they identify as~: $\psi_- \equiv \psi (4040)$ and $\psi_+
\equiv \psi (4160)$, $H_C$ is an hybrid state. They use the selection
rule for hybrid decay which, in the flux tube model, holds for the
lowest lying hybrid as already mentioned \cite{isgur, page}.

In order to  study the possibility for the OCP scenario in the present model
we will compute the mixing pattern of the conventional $\psi
(3S)$ with the hybrid charmonium $c\bar{c}g, 1^{--}$ mesons, listed in table 1.
~From the results about the decay widths, only the gluon-excited one, $l_g=1$
might fit into the OCP scenario, since, in the absence of mixing with the
quark-excited ones it verifies the same decay selection rule as in the
flux-tube model. For the sake of completeness, we also compute the mixing with
the quark-excited hybrids.

\subsection{Hybrid-charmonium transition Hamiltonian} \hspace*{\parindent}

The transition Hamiltonian is given by QCD ~:

\beq
H_{trans} = g \int dx \bar{\psi} (x) \ \gamma_{\mu} \ {\lambda^a \over 2} \ \psi 
(x) \ A_a^{\mu}(x)
\label{39f}
\eeq

\noi from (\ref{16e}), (\ref{17e}) and (\ref{18e}) we express the Hamiltonian 
in Fock space and expand to the first non vanishing order in $v/c$ (i.e. the
first relativistic correction) and get
$H_{trans} = H_1 + H_2$ where $H_1$ ($H_2$) is the Hamiltonian in 
which the quark (antiquark) emits a gluon, with 

\bea \label{40e} &&H_1 = g \sum_{\lambda \lambda ' \bar{\lambda}
\bar{\lambda '}\atop{ab'b \bar{b}\bar{b}'}} \int {d^3p_c \ d^3p_{c'}\
d^3k \over (2 \pi )^9 \ \sqrt{2 \omega}} \ (2 \pi)^3 \ \delta \left (
p_c - k - p_{c'} \right ) b^+_{p_{c}' \lambda ' b'} \ b_{p_c \lambda b}
{\lambda^a_{b'b} \over 2} \ a^+_{k\mu_g a} \nn \\ &&\delta_{\bar{b},
\bar{b}'} \delta_{\bar{\lambda}, \bar{\lambda} '} \chi^+_{\lambda '}
\left [ {\vec{\varepsilon}^{(\mu_g)} \left ( \vec{p}_c + \vec{p}_{c'}
\right ) \over 2m} + {\vec{\varepsilon}^{(\mu_g)} \vec{k} \over m_g} +
{i \vec{\sigma} \over 2m} (\vec{\varepsilon}^{(\mu_g)} \wedge \vec{k})
\right ] \chi_{\lambda} \ dp_{\bar{c}} \ \delta _3\left ( p_{\bar{c}} -
p_{\bar{c}'} \right ) \eea

\noi where $\lambda \lambda '$ are the quark spin properties, $b,b'$ the quark 
color indices and
$a$ labels the gluon colors, $\vec{\varepsilon}^{(\mu_g)}$ the gluon 
polarisation and~:

\bea \label{41e} &&H_2 = - g \sum_{\lambda \lambda ' \bar{\lambda}
\bar{\lambda '}\atop{ab'b \bar{b}\bar{b}'}} \int {d^3p_{\bar{c}} \
d^3p_{\bar{c}'}\ d^3k \over (2 \pi )^9 \ \sqrt{2 \omega}} \ (2 \pi)^3 \
\delta \left ( p_{\bar{c}'} + k - p_{\bar{c}} \right )
d^+_{p_{\bar{c}'}\bar{\lambda} ' \bar{b}'} \ d_{p_{\bar{c}}\bar{\lambda}
\bar{b}} {\lambda^a_{\bar{b}'\bar{b}} \over 2} \ a^+_{k\lambda a} \nn \\
&&\delta_{b, b'} \delta_{\lambda, \lambda '} \chi^+_{\bar{\lambda} '}
\left [ {\vec{\varepsilon}^{(\mu_g)} \left ( \vec{p}_{\bar{c}} +
\vec{p}_{\bar{c}'} \right ) \over 2m} + {\vec{\varepsilon}^{(\mu_g)}
\vec{k} \over m_g} + {i \vec{\sigma} \over 2m}
(\vec{\varepsilon}^{(\mu_g)} \wedge \vec{k}) \right ]
\chi_{\bar{\lambda}} \ dp_{c} \ \delta_3 \left ( p_{c} - p_{c'} \right )
\ . \eea

\noi The (3S) charmonium meson is given in the non-relativistic approximation 
by~:

\bea
\label{42e}
|\psi(3S)> &=& \sum_{\lambda \lambda ' m_{c \bar{c}} a } \int {d^3p_c \
d^3p_{\bar{c}} \over (2 \pi )^6} \ {1 \over \sqrt{3}} \ \chi^{\lambda_{c\bar 
c}}_{\lambda 
\lambda '} (2 \pi )^3 \
\delta_3 \left ( p_c + p_{\bar{c}} \right ) \nn \\
&&\psi_{c\bar{c}} \left ( {p_c - p_{\bar{c}} \over 2} \right ) <\ell_{c 
\bar{c}}, m_{c \bar{c}} S_{c\bar c}
\lambda_{c \bar{c}} |J_{c\bar{c}}  M_{c \bar{c}}> \ b^+_{p_c\lambda a} \ 
d^+_{p_{\bar{c}} \lambda'
 {a}} 
\eea

\noi where~: 
\beq
\psi_{c\bar{c}}(p_c) = \left ( {16 \pi^3 R_{c\bar{c}}^3 \over \Gamma (3/2)} 
\right)^{1/2} \ {4
\over \sqrt{30}} \ L_2^{1/2}(R^2_{c\bar{c}} p_c^2) y_{\ell_{c 
\bar{c}}}^{m_{c\bar{c}}}(\vec{p}_c )\,\exp \left [ -
{1 \over 2} \left ( R^2_{c\bar{c}} p_c^2 \right ) \right ] \ . \label{43e}
\eeq

\noi $L^{1/2}_2(x)$ being the Laguerre polynomial of the second order, since the 
3S
charmonium is a second radial excitation wave function.

To compute the charmonium-hybrid transition we will calculate the matrix 
element:
\beq
I^{\ell_{q\bar{q}}(\ell_g)} \equiv <c\bar{c} g (hybrid)|H_{trans}|\psi(3S)>.
\eeq

\noi We obtain~:

\bea \label{44e} I^{{\ell_{c'\bar{c}'}({\ell_g})}} &=& 2Bg \sum_{\lambda
, \lambda ', m_g, m_{c\bar{c}}, m_{c'\bar{c}'}\atop{\mu_g , \lambda ,
\bar{\lambda}, \bar{\lambda}'}} \int {d^3p_c \ d^3p_c' \over (2 \pi)^6
\sqrt{2 \omega}} \ {2 \over \sqrt{3}} \nn \\
 & & \chi^+_{\lambda '} \left [ \vec{\varepsilon}^{(\mu_g)} {\left (
\vec{p}_c + \vec{p}_{c'} \right ) \over 2m} +
\bar{\varepsilon}^{(\mu_g)} {\left ( \vec{p}_c - \vec{p}_{c'} \right )
\over m_g} + {i \vec{\sigma} \over 2m} \left (
\vec{\varepsilon}^{(\mu_g)} \wedge (\vec{p}_c - \vec{p}_{c'} \right )
\right ] \chi_{\lambda} \nn \\
&&\psi^{m_{c'\bar{c}'}*}_{\ell_{c'\bar{c}'}} \left ( {\vec{p}_c +
\vec{p}_{c'} \over 2} \right ) \psi_{\ell_g}^{m_g*} \left ( \vec{p}_c -
\vec{p}_{c'} \right ) \psi_{\ell_{c\bar{c}}}^{m_{c\bar{c}}} (\vec{p}_c)
\ <\ell_{c\bar{c}} \ m_{c\bar{c}} \ S_{c \bar{c}} \ \lambda_{c\bar{c}}
|J_{c\bar{c}} \ M_{c\bar{c}}> \nn \\ &&<\ell_g \ m_g \ 1 \ \mu_g|J_g \
M_g> \ <\ell_{c'\bar{c}'} \ m_{c'\bar{c}'} \ J_g \ M_g|Lm> \ <L \ m \
S_{c'\bar{c}'} \ \lambda_{c'\bar{c}'}|JM> \nn \\ &&<{1 \over 2} \
\lambda \ {1 \over 2} \ \bar{\lambda}|S_{c\bar{c}} \ \lambda_{c\bar{c}}>
\ < {1 \over 2} \ \lambda ' \ {1 \over 2} \
\bar{\lambda}'|S_{c'\bar{c}'} \ \lambda_{c'\bar{c}'}> \ .
 \eea 
\noi Where~:
\bea
\label{44f}
B  &=&  ( {{16 \pi^3})^{3 / 2}}{R_{c'\bar{c}'}^{5/2}} \ R_g^{3/2} \ 
R_{c\bar{c}}^{3/2}
\over \pi \sqrt{30(\Gamma(3/2))^2  \Gamma(5/2)}
\eea

\noi We have used in the calculation generating function $w(x, t)$ of the 
Laguerre polynomials, which
is easier to handle than the Laguerre polynomials themselves~:

\beq
\label{45e}
w(x,t) = (1 - t)^{-a-1} \ e^{-xt/(1 - t)} = \sum_{x=0}^{\infty} L_x^{\alpha} (x) 
\ t^n \quad ,
\quad |t| < 1 \quad .
\eeq 

\noi From (\ref{45e}) and (\ref{44e}) we obtain~:

\bea
\label{46e}
I^{\ell_{c'\bar{c}'}=1} &=& 2 \alpha_s \Big [ \left ( {1 \over 2m} T^{1} + {1 
\over m_g} \
T^{2} \right ) 3 \delta_{S_{c\bar{c}}, S_{c'\bar{c}'}} \ \delta_{L,0} \ 
\delta_{J,J_{c\bar{c}}} \
\delta_{M,M_{c\bar{c}}} \nn \\
&&+ {i \over 2m} \ T^{2} (-2i) \ \delta _{J,J_{c\bar{c}}} \ 
\delta_{M,M_{c\bar{c}}}
\delta_{L,S_{\sigma}} \Big ]
\eea

\noi and

\beq
\label{47e}
I^{\ell_g = 1} \ = \ 2 \alpha_s \ {i \over 2m} \ T^{4} (-2i) \delta 
_{J,J_{c\bar{c}}} 
\ \delta_{M,M_{c\bar{c}}}
\delta_{L,S_{\sigma}} 
\eeq

\noi where~:

\bea
&&T^1 = - 0.0009446 \ \hbox{GeV}^2 \quad , \nn \\
&&T^2 = - 0.0005347 \ \hbox{GeV}^2 \quad , \nn \\
&&T^3 = - 0.0003679 \ \hbox{GeV}^2 \quad , \nn \\
&&T^4 = - 0.0007595 \ \hbox{GeV}^2 \quad . 
\eea

Finally the transition amplitudes between the 3S charmonium and the hybrid meson 
are listed in
table 5 :

\begin{table}[htb]
\begin{center}
\begin{tabular}{|c|c|c|c|c|c|c|c|}
\hline
& & & & & & & \\
$P$ &$C$  &$\ell_g$ &$\ell_{c'\bar{c}'}$ &$J_g$ &$S_{c'\bar{c}'}$ &$L$
&$<c\bar{c}g|H_{tr}|c\bar{c}>$ ($\alpha_s$Mev)  \\
& & & & & & & \\  
\hline
$-$ &$-$ &$0$ &$1$ &$1$ &$1$ &$0$ &$- 5.6$ \\
\hline
$-$ &$-$ &$0$ &$1$ &$1$ &$1$ &$1$ &$-0.629$ \\
\hline
$-$ &$-$ &$0$ &$1$ &$1$ &$1$ &$2$ &$0$ \\
\hline
$-$ &$-$ &$1$ &$0$ &$1$ &$0$ &$1$ &$-0.89$ \\
\hline
\end{tabular}
\end{center}
\caption{\leftskip 1pc \rightskip 1pc
\baselineskip=10pt plus 2pt minus 0pt Predicted transition amplitudes between 
the 3S charmonium and the
hybrid mesons in MeV times $\alpha_s$.}
\label{tab5}\vskip 5mm
\end{table}

\noi These matrix elements turn out to be very small, of the order of 1 MeV. 
\noi Using these transition matrix elements and the fact that $E_+=4160$ Mev 
and$ E_-=4040 $Mev are the expected 
eigenvalues of
the transition Hamiltonian (\ref{39f}), it is easy to  compute the mixing angle
between the (3S) charmonium and the $c\bar{c}g$ hybrid meson:
\beq
\label{49e}
\tan(\theta) = {2 |H_{12}|\over{\sqrt{ (E_{+}-E_{-})^2 -4|H_{12}|^2}}}
\eeq
\noi where $H_{12}$ is one of the transition matrix  elements given in table 5. 
The result are
listed in table 6~:
\begin{table}[htb]
\begin{center}
\begin{tabular}{|c|c|c|c|c|c|c|c|}
\hline
& & & & & & & \\
$\ell_g$ &$J_g$  &$\ell_{c'\bar{c}'}$ &$S_{c'\bar{c}'}$ &$L$ &$H_{12}(Mev)$ 
&$\tan(\theta)$
&$\theta$   \\
& & & & & & & \\  
\hline
$0$ &$1$ &$1$ &$1$ &$0$ &$5.6$ &$0.0937$ &$0.0934$ \\
\hline
$0$ &$1$ &$1$ &$1$ &$1$ &$0.629$ &$0.0105$ &$0.0105$ \\
\hline
$0$ &$1$ &$1$ &$1$ &$2$ &$0$ &$0$ &$0$ \\
\hline
$1$ &$1$ &$0$ &$0$ &$1$ &$0.89$ &$0.0148$ &$0.0148$ \\
\hline
\end{tabular}
\end{center}
\caption{\leftskip 1pc \rightskip 1pc
\baselineskip=10pt plus 2pt minus 0pt Predicted mixing angles between the (3S) 
charmonium and the $c\bar{c}g$ hybrid meson}
\label{tab6}\vskip 5mm
\end{table}

\noi

These mixing angles turn out to be very small, because the transition matrix 
elements are much smaller than the level spacing between $\psi(4040)$ and 
$\psi(4160)$. 
This feature invalidates the  the (OCP) scenario which needs a rather large 
mixing angle 
as exemplified in eq. (\ref{37e}). A small mixing angle leaves us with two 
states 
very similar to the starting ones: one hybrid, weakly coupled with the photon 
and thus incompatible with any of the two $\psi(4040)$ and $\psi(4160)$ states, 
and only one $\psi(3S)$ whcih cannot account for the latter two states.

The smallness of the transition matrix elements is due to the 
quasi-orthogonality between the conventional $\psi
(3S)$ and the wave functions of the $c\bar c$ pair in  the hybrid, namely 
$(1S)$ or $(1P)$ state. The recoil of the gluon and the difference of the radii 
is not enough to eliminate 
this quasi-orthogonality
suppression. As a check 
 we have also  computed 
 the matrix elements between the 
$(1S)$ charmonium and the hybrids, and we obtain  
mixing amplitudes which are two orders of magnitude larger than with the $(3S)$.

\section{Conclusion}\hspace*{\parindent}

	We have considered  the  $J^{PC}=1^{--}$ hybrid $c \bar c g$ mesons
	in a quark and gluon constituant model. Using for simplicity a 
chromoharmonic potential (\ref{1e}\ref{2e}), we 
have computed the spectrum and wave functions of these hybrids, their decay 
widths into 
ground state mesons and finally their mixing with $\psi(3S)$. 
	
Restricting ourselves to the lightest states ($l_g+l_{c\bar c}$=1)
 we have found 
one hybrid meson with an 
orbital excitation between  the gluon and the $c\bar c$ system ($l_g=1$), and 
three with an orbital 
excitation between the $c$ and the $\bar c$ ($l_{c\bar c}=1$). The former is 
stable for decay into ground 
state meson (the only opened channels around 4.1 GeV). The latter have so large 
widths that they do not 
really exist as resonances. What happens here is simply that nothing prevents 
the constituant gluon to decay instantaneously into a quark pair.
Within the chromoharmonic model, these two classes do not mix, leaving alive 
only one hybrid meson  ($l_g=1$) which 
does not decay. This patter of two types of hybrids, one forbidden, teh other allowed to decay into ground state mesons is quite similar to the results fo the flux tube model, except for a reverse ordering of masses. We do not know what decay width the latter model would predict for the allowed decay.

We have thus considered the latter as a candidate for the OCP scenario which 
explains anomalies in the pair
$\psi(4040)-\psi(4160)$ as the result of a large-angle mixing between a hybrid 
meson and the $\psi(3S)$ 
charmonium. 
However, the very small mixing Hamitonian matrix elements (due to the
quasi-orthogonality of the $(3S)$\, ${c\bar {c}}$ 
wave function in the charmonium and the $(1S)$ or $(1P)$ in the hybrid) and the 
very small resulting angles lead us to exclude 
the OCP scenario in view of the observed mass splitting  $\psi(4040)-\psi(4160)$.

Finally it should be noted that a more realistic potential might induce a large 
mixing between the two classes 
of hybrids, resulting in very large widths for all hybrids and finally no real 
resonant hybrid.  Only a small 
attraction in the two meson continuum due to a very short-lived hybrid ``state'' might then remain. This might look like what has been seen in the analogous 
light meson $1^{-+}$ sector where a possibly hybrid induced  ``activity''  may
have been seen, as 
noted in \cite{faessler}.

If such a large mixing between the different classes of hybrids does not exist,
the search for $l_g=1$ hybrids remains  an important task, although made 
difficult by the absence of coupling 
to the ground state mesons.

Anyhow a new explanation for the  $\psi(4040)-\psi(4160)$ mystery has still to be found in view of the difficulties 
of the OCP scenario.

\section*{Acknowledgements}

\hspace*{\parindent} It is a pleasure to thank Suh-Urk Chung 
 for valuable discussions. We are particularly indebted to Philip Page
 for very useful comments and for a careful and critical reading of 
 the draft.

\end{document}